\pdfoutput=1
\documentclass[aps,prd,preprint,superscriptaddress,showpacs,preprintnumbers]{revtex4-1}
\usepackage[colorlinks=true, pdfstartview=FitV, linkcolor=red, citecolor=blue, urlcolor=black, pdftitle={Temporal Chiral Spiral in Strong Magnetic Fields},pdfauthor={Tomoya Hayata, Yoshimasa Hidaka, Arata Yamamoto},pdfsubject={}, pdfkeywords={}]{hyperref}

\usepackage[dvipdfmx]{graphicx}
\usepackage{amssymb,bm}
\usepackage{amsthm,amsmath}
\usepackage{color}
\newcommand{\Sl}[1]{{\ooalign{\hfil/\hfil\crcr$#1$}}}
\newcommand{\SlD}{\Sl{D}}
\newcommand{\SlDr}{\Sl{D}^{(r)}}
\newcommand{\NL}{N_L}
\newcommand{\Nc}{N_c}
\newcommand{\cz}{\xi}
\newcommand{\tr}{{\rm tr}}
\newcommand{\Tr}{{\rm Tr}}

\begin{document}
\preprint{RIKEN-QHP-92}
\title{Temporal Chiral Spiral in Strong Magnetic Fields}
\author{Tomoya Hayata}
\affiliation{Department of Physics, The University of Tokyo, Tokyo 113-0031, Japan}
\affiliation{Theoretical Research Division, Nishina Center, RIKEN, Wako 351-0198, Japan}
\author{Yoshimasa Hidaka}
\affiliation{Theoretical Research Division, Nishina Center, RIKEN, Wako 351-0198, Japan}
\author{Arata Yamamoto}
\affiliation{Theoretical Research Division, Nishina Center, RIKEN, Wako 351-0198, Japan}

\date{\today}

\begin{abstract}

 Vacuum properties of quantum chromodynamics in strong magnetic and finite electric fields are investigated.
 We show that when a uniform electric field is instantaneously applied in the parallel direction to a strong magnetic field, 
 it induces temporal oscillation of the scalar and pseudoscalar condensates.
 This is a temporal analog to the chiral spiral.
 The oscillation originates with the propagation of the collective mode, which is protected by the axial anomaly and thus is nondissipative.
 
\end{abstract}
\maketitle

 \section{Introduction}
 Inhomogeneous order parameters are ubiquitous in nature. 
 Famous examples are charge- and spin-density waves in condensed matter physics.
 For example, in $(1+1)$ dimensions,
 left- and right-moving electrons form a standing wave
 with an oscillation of the charge and spin densities 
 accompanied by a periodic distortion of the underlying lattice~\cite{Peierls:1955,Overhauser:1960}.
 One of the most intriguing phenomena of density waves is the ``sliding'' of them
 by applying electric or magnetic fields, 
 whose effect on transport properties of systems has been actively investigated~\cite{Gruner:1988}.

 A similar periodic order parameter in quantum chromodynamics (QCD)
 has been investigated in the context of chiral symmetry breaking.
 It was originally discussed in the QCD matter at high density 
 with a large number of colors~\cite{Deryagin:1992rw}.
 The scalar and pseudoscalar condensates locally form
 and vary with the spatial position in a spiral form, which is called the ``chiral spiral'' \cite{Park:1999bz,Kojo:2009ha}.
 This is an analog to the spin-density wave~\cite{Overhauser:1960}.
The chiral spiral was also proposed in strong magnetic fields~\cite{Basar:2010zd},
 which is closely related to the chiral magnetic effect~\cite{Kharzeev:2007jp,Fukushima:2012vr}
 and the chiral magnetic wave~\cite{Kharzeev:2010gd}.
 The charge-dependent azimuthal correlation, which is a key observable of the chiral magnetic effect, has been recently observed
 in the relativistic heavy-ion collision experiments at RHIC~\cite{Abelev:2009ad}.

 While a spatially inhomogeneous state is widely known, 
 a temporally inhomogeneous state is quite nontrivial. 
 In a temporally inhomogeneous state, time translational symmetry is broken. 
 The order parameter continues to change in time evolution. 
 The spontaneous generation of temporal inhomogeneity is a great challenge in theoretical physics \cite{Wilczek:2012jt}.

\begin{figure}
\includegraphics[scale=.40]{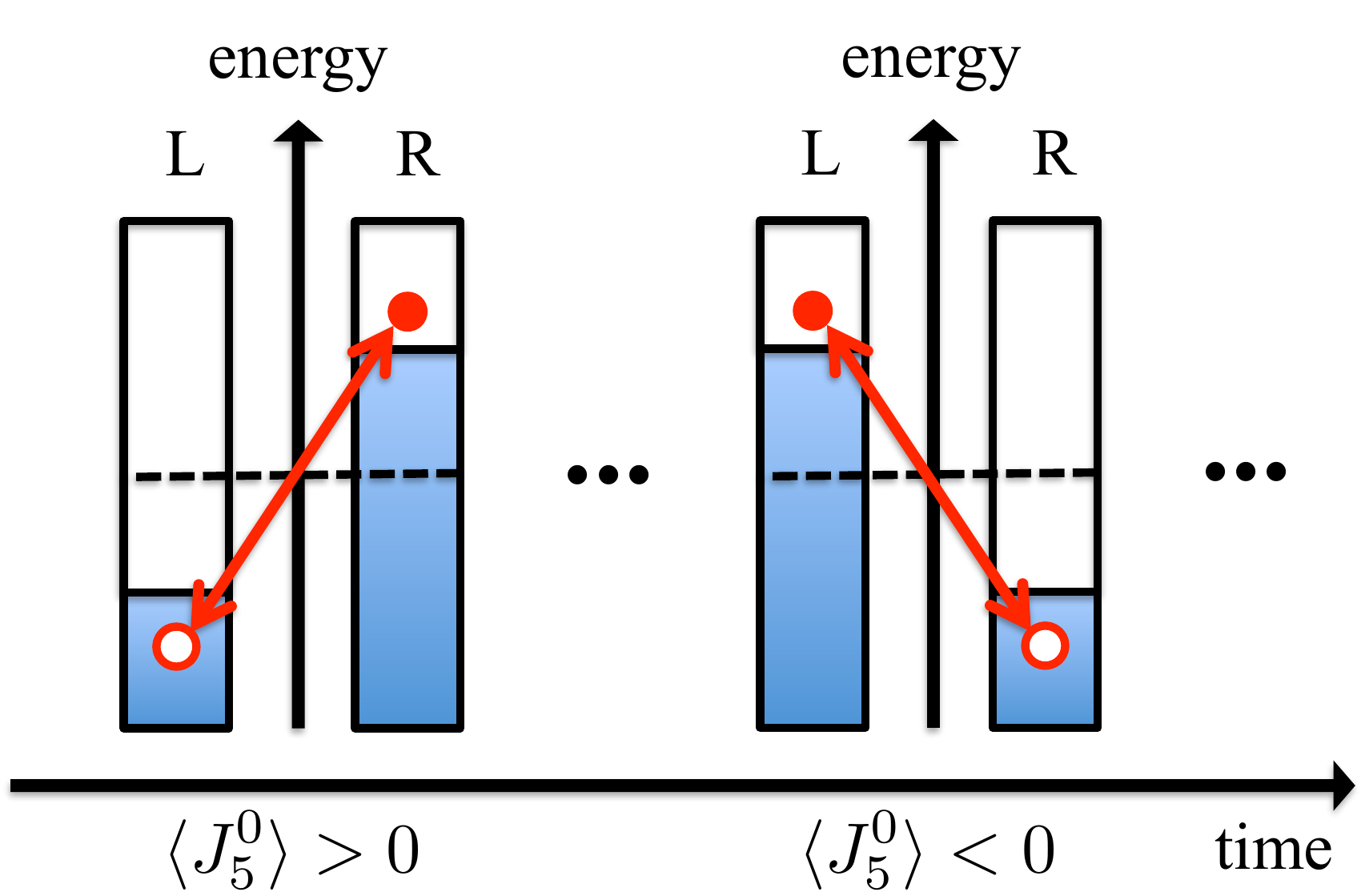}
\caption{\label{fig1}
Energy eigenstates of massless quarks with right- and left-handed chirality 
in an oscillating axial charge $\langle J_5^0\rangle$.
The blue regions indicate occupied eigenstates.
The red dots and open circles denote the particle and antiparticle excitations, respectively.
}
\end{figure}

 The main goal of this paper is to demonstrate the formation of time-periodic scalar and pseudoscalar condensates with a spiral structure,
 which is a temporal analog to the spatial chiral spiral.
 Our idea is based on two intriguing phenomena of QCD at finite electric and magnetic fields:
 (I) quark and antiquark pairs creation by external electric fields in the QCD vacuum,
 which is known as the Schwinger mechanism~\cite{Schwinger:1951nm}.
 The backreaction from quark pairs
 induces the temporal oscillation of electric current~\cite{Tanji,Iwazaki}.
 (II) At strong magnetic fields, the mixing of the electric current  parallel to the magnetic field $J^{3}$ and the axial charge $J_5^0$ occurs through the axial anomaly,
 which is known as the chiral magnetic effect~\cite{Kharzeev:2007jp}.
 Because of (I) and (II), an external electric field instantaneously applied parallel to a strong magnetic field
 generates a temporal oscillation of $\langle J_5^0\rangle$.  
 In the QCD vacuum with the oscillating $\langle J_5^0\rangle$,
 the scalar condensate becomes time dependent as schematically depicted in Fig.~\ref{fig1}.
 For massless quarks, at $\langle J_5^0\rangle>0$, right-handed particle ($R+$) and left-handed antiparticle states ($L-$) are occupied as in Fig.~\ref{fig1}.
In such a case, the scalar condensate is formed by
 the pairing of a right-handed quark with a left-handed antiquark $\langle q^{\dagger}_{L-}q_{R+}\rangle$.
 On the other hand, at $\langle J_5^0\rangle<0$,
 the pairing of a left-handed quark ($L+$) with a right-handed antiquark ($R-$)
 $\langle q_{R-}^{\dagger}q_{L+}\rangle$ forms the scalar condensate.
 Then, if $\langle J_5^0\rangle$ changes in time,
 the scalar condensate is formed by the interplay of these pairings, and becomes time dependent.
 We will show that the scalar and pseudoscalar condensates oscillate in a spiral form, and
 call it the ``temporal chiral spiral.'' 

 In this paper, we explicitly derive the temporal chiral spiral in the lowest Landau level (LLL) approximation.
 Since quarks in the LLL approximation are reduced to the $(1+1)$-dimensional theory, 
 we can solve real-time dynamics of the scalar and pseudoscalar condensates 
 in electric fields at the full quantum level.
 We note, however, that the above mechanism may work in the presence of higher Landau levels as well.
 This paper is organized as follows: In Sec.~\ref{sec:dimensional reduction}, we summarize the standard method 
 to study relativistic fermions in strong magnetic fields, that is, the LLL approximation.
 We show that the dynamics of relativistic fermions are reduced to that of the $(1+1)$-dimensional theory in strong magnetic fields, 
 and derive the effective action of QCD in the strong-$B$ limit. Then, in  Sec.~\ref{sec:nonabelian bosonization},
 we rewrite this effective action in terms of bosonic degrees of freedom, 
 which is useful to study nonperturbatively the effect of electric fields on the scalar and pseudoscalar condensates.
 In  Sec.~\ref{sec:temporal chiral spiral}, we show how the scalar and pseudoscalar condensates behave 
 when external electric fields are applied parallel to strong magnetic fields.
 In particular, we consider two cases of the external electric fields: (I) the instantaneous electric field $E(t)=(E_0/m_{\gamma})\delta(t)$;
 (II) the constant electric field $E(t)=E_0\Theta(t)$ [$\Theta(t)$ is the Heaviside step function].
 After that, we discuss the relation between the temporal chiral spiral and the chiral magnetic wave, 
 and show that the nondissipative nature of the temporal chiral spiral originates with the axial anomaly in $(1+3)$ dimensions.
 We also investigate the effect of finite densities and propose a new nonequilibrium stationary state in a dense quark matter. 
 Section~\ref{sec:concluding remarks} is devoted to concluding remarks.
In Appendixes \ref{sec: propagator} and \ref{sec: nonabelian bosonization}, the detailed derivation of the Landau quantization for fermions  and the non-abelian bosonization technique are shown.

%---------------------------------------------
%-----------------------

 \section{Dimensional reduction}
 \label{sec:dimensional reduction}
{We start with the dimensional reduction of quarks in strong magnetic fields.
Let us consider a magnetic field $B$ applied in the $z$ direction.
 (For late use, we take $e>0$ and $B>0$.)
The energy spectra of relativistic fermions and antifermions are discretized by the Landau quantization~\cite{Landau},
 \begin{equation}
\varepsilon(n,{p^3},s)=\pm\sqrt{(p^3)^2+2eB(n+1/2\mp s)},
\label{Landau level}
\end{equation}
with $n$ and $s=\pm1/2$ denoting the Landau level and the spin, respectively (For the details of the Landau quantization,
see Appendix \ref{sec: propagator}.)
 The energy eigenstate with $n>0$ has the induced mass $m^2_n=2eBn$.
 When the magnetic field is much larger than the QCD energy scale, namely, $eB\gg {\Lambda_{\rm QCD}}^2 \sim (200 \, {\rm MeV})^2$,
 the contributions from higher Landau levels ($n>0$) are suppressed due to the induced mass, and the LLL ($n=0$ and $s=1/2$) becomes dominant.
Thus, the quark field can be projected to the LLL states as~\cite{Hidaka:2011dp}
 \begin{equation}
 q_{\rm LLL}(x)=
 \begin{pmatrix}
  \sum_{l=0}^{\NL}\varphi_l(x_\perp)c_f(x_{\parallel}) \\
  0
 \end{pmatrix}
 \label{LLLprojection}
 \end{equation}
 with $f=(l,i)$, where $l$ and $i$ denote the angular momentum and the color, respectively.
 Here, $\varphi_l(x_\perp)$ is the wave function of the LLL state in $x_{\perp}=(x^1,x^2)=(x,y)$ shown in Eq.~(\ref{LLL wave function}), 
and $c_f(x_{\parallel})$ is the two-component Dirac field in $x_{\parallel}=(x^0,x^3)=(t,z)$.
 The degeneracy of the LLL states is given by $\NL=eBV_{\perp}/2\pi$ with the transverse volume $V_{\perp}$. }

Substituting Eq.~(\ref{LLLprojection}) into the quark action coupled to 
 the dynamical U($1$) and SU($N_c$) gauge fields, $A^{\text{em}}_\mu$ and $A^a_\mu$, in $(1+3)$ dimensions,
it is reduced to the $(1+1)$-dimensional quark action~\cite{Gusynin:1995nb,Hidaka:2011dp,Kojo:2012js},
\begin{equation}
  S_{\mathrm{LLL}}= \int d^2x_{\parallel}\;\Bigl(i\bar{c}_f\gamma^{\mu}(\partial_{\mu}+iea_{\mu})c_f
    +\bar{c}_f\gamma^{\mu}[A_{\mu}]_{ff^{\prime}}c_{f^{\prime}}\Bigr) .
 \label{LLLaction}
\end{equation}
In the following, $\gamma^\mu$ represent the gamma matrices in $(1+1)$ dimensions, not in $(1+3)$ dimensions, unless otherwise stated.
For late use, we explicitly write down  the gamma matrices in the chiral basis:
\begin{align}
 \gamma^0=
 \begin{pmatrix}
  0 & 1 \\
  1 & 0
 \end{pmatrix}  &, \quad
\gamma^3=
 \begin{pmatrix}
  0 & -1 \\
  1 & 0
 \end{pmatrix} , \\
 \gamma^+=\gamma_-\equiv\gamma^0+\gamma^3=
 \begin{pmatrix}
  0 & 0 \\
  2 & 0
 \end{pmatrix} &, \quad
\gamma^-=\gamma_+\equiv\gamma^0-\gamma^3=
 \begin{pmatrix}
  0 & 2 \\
  0 & 0
 \end{pmatrix}.
\end{align}
{The $(1+1)$-dimensional ``photon'' and ``gluon'' fields are constructed from the parallel components of the $(1+3)$-dimensional photon and gluon fields, $A^{\text{em}}_\mu$ and $A^a_\mu$ ($\mu = 0,3$) \cite{Hidaka:2011dp}.
The $(1+1)$-dimensional ``photon'' field $a_{\mu}$ is composed of the photon field $A^{\text{em}}_\mu$ with zero transverse momentum,
\begin{equation}
{a}_{\mu}(x_{\parallel})\equiv\int d^2x_\perp A^{\text{em}}_{\mu}/V_\perp.
\end{equation}
The $(1+1)$-dimensional ``gluon'' field $A_{\mu}$ is composed of the photon field $A^{\text{em}}_\mu$ with nonzero transverse momentum and the gluon field $A^a_\mu$ with both zero and nonzero transverse momenta,
\begin{equation}
[A_{\mu}]_{ff'}(x_{\parallel})\equiv\int d^2x_\perp 
 \Bigl(-eA^{\text{em}}_{\mu}\delta_{ii'}+gA^a_{\mu}t^a_{ii'}\Bigr)\varphi_{l}^{*}(x_\perp)\varphi_{l'}(x_\perp)+ea_\mu \delta_{ii'}\delta_{ll'}.
\label{gluon}
\end{equation}
The perpendicular components of the $(1+3)$-dimensional photon and gluon fields, $A^{\text{em}}_\mu$ and $A^a_\mu$ ($\mu=1,2$), decouple from the dimensionally reduced quarks.
 We remark that the perpendicular dynamics of quarks is not completely lost even in the LLL approximation.
 The perpendicular degrees of freedom of quarks are mapped into the internal ones (the angular momentum) of the dimensionally reduced quarks.
They couple to the ``gluon'' fields with nonzero transverse momentum.}

 \section{Non-abelian bosonization}
 \label{sec:nonabelian bosonization}
 Now, we rewrite the action (\ref{LLLaction}) by using the bosonization techniques~\cite{Frishman:1992mr}.
(For the details of the derivation, see Appendix \ref{sec: nonabelian bosonization}.)
In $(1+1)$ dimensions, vector fields can be decomposed as 
\begin{align}
-ea_{\mu}&=\frac{i}{2}(g_{\mu\nu}-\epsilon_{\mu\nu})v\partial^{\nu}v^{-1}+\frac{i}{2}(g_{\mu\nu}+\epsilon_{\mu\nu})u^{-1}\partial^{\nu}u,
\label{U(1)}\\
A_{\mu}&=\frac{i}{2}(g_{\mu\nu}-\epsilon_{\mu\nu})V\partial^{\nu}V^{-1}+\frac{i}{2}(g_{\mu\nu}+\epsilon_{\mu\nu})U^{-1}\partial^{\nu}U,
\label{SU(NLNc)}
\end{align}
where $\epsilon^{\mu\nu}$ are totally antisymmetric tensors with $\epsilon^{03}=1$.
With the combinations of the local vector and axial-vector gauge transformations
\begin{equation}
 \Omega=
 \begin{pmatrix}
  vV & 0 \\
  0 & u^{-1}U^{-1}
 \end{pmatrix} , \quad
\overline{\Omega}=
 \begin{pmatrix}
  uU & 0 \\
  0 & v^{-1}V^{-1}
 \end{pmatrix} ,
\label{Omega}
\end{equation}
the action~(\ref{LLLaction}) is rewritten as 
\begin{equation}
  S_{\mathrm{LLL}}= \int d^2x_{\parallel}\; \bar{c} \Omega i\gamma^{\mu}\partial_{\mu}\overline{\Omega} c.
\label{Dirac operator}
\end{equation}
By using the transformation $c^{\prime}=\overline{\Omega} c$, we can remove $a_{\mu}$ and $A_{\mu}$, and transform the action to the free quark action. 
The resultant free quark action only gives an irrelevant normalization factor to the generating functional $W[a_{\mu},A_{\mu}]$.
However, since the measure of quark fields is not trivially transformed under the axial-vector transformation,
the Jacobian gives an anomalous contribution to $W[a_{\mu},A_{\mu}]$.
This anomalous contribution can be calculated by the Fujikawa method~\cite{Fujikawa:1979ay,Frishman:1992mr} and results in the so-called Wess-Zumino-Witten action
\begin{align}
\Gamma[G]&=\frac{1}{8\pi}\int d^2x_{\parallel}\tr\left[\partial^{\mu}G^{-1}\partial_{\mu}G\right]\notag\\
&\quad-\frac{1}{4\pi}\int_0^1 dr\int d^2x_{\parallel}\;\epsilon^{\mu\nu}\tr\left[\left(G_r^{-1}\partial_rG_r\right)\left(G_r^{-1}\partial_\mu G_r\right)\left(G_r^{-1}\partial_\nu G_r\right)\right] ,
\label{WZW action}
\end{align}
where the trace runs over the ``color" space.
 The generating functional is given as~\cite{Frishman:1992mr} (see Appendix \ref{sec: nonabelian bosonization})
\begin{align}
  W[a_{\mu},A_{\mu}] &= W_1[a_{\mu}]+W_2[A_{\mu}]
  \notag \\
  &=-\Gamma[uv]-\Gamma[UV] .
\label{generating functional}
\end{align}
Introducing two boson fields, a U($1$) generator $\phi$, and an SU($\NL \Nc$) element $h$, we can write the bosonic action
\begin{equation}
  S_{\rm LLL} =\int d^2x_{\parallel}\;\Bigl(\frac{1}{2}\partial^{\mu}\phi\partial_{\mu}\phi-ej^{\mu}a_{\mu}\Bigr)
  +\Gamma[h]+\int d^2x_{\parallel}\tr\left[k^{\mu}A_{\mu}-\frac{1}{4\pi}\left(A^{\mu}A_{\mu}-h^{-1}A_{+}hA_{-}\right)\right],
  \label{BosonizedAction}
\end{equation}
where $A_{\pm}=A_0\pm A_3$.
 $j^{\mu}(\phi)$ and $k^{\mu}(h,A_{\mu})$ are the electric and ``color'' covariant currents and are defined as 
\begin{align}
  j^{\mu} &\equiv
  \sqrt{\frac{\NL \Nc}{\pi}}\epsilon^{\mu\nu}\partial_{\nu}\phi,
\\
  k_{+} &\equiv
  \frac{i}{2\pi}\left(U^{-1}\partial_+U-(Uh)^{-1}\partial_+(Uh)\right),
\\
 k_{-} &\equiv
\frac{i}{2\pi}\left(V\partial_+V^{-1}-(hV)\partial_+(hV)^{-1}\right).
\end{align}
The first term comes from the abelian part $W_1[a_{\mu}]$ and the second and third terms come from the non-abelian part $W_2[A_{\mu}]$.
We remark that $\phi$ couples only to the ``photon'' fields $a_{\mu}$. 
In addition, the ``photon'' fields $a_{\mu}$ do not couple to any other components of gauge fields.
Therefore, $\phi$ and $a_\mu$ are not affected by perpendicular or color dynamics of $h$ and the ``gluon'' fields $A_{\mu}$.
We only have to solve the abelian part of the action to study the real-time dynamics of $\phi$ and $a_\mu$.

The quickest derivation of Eq.~(\ref{BosonizedAction}) is to utilize the invariance of the path integral measure.
For the abelian part, if we define the bosonic action of an element of U($1$) $\Phi$ as 
\begin{equation}
\widetilde{S}_{\rm LLL}(\Phi,a)\equiv \Gamma[u\Phi v]-\Gamma[uv] ,
\end{equation}
we can show that
\begin{equation}
iW_1[a_{\mu}]=\ln\left(\int {\cal D}\Phi\;e^{i\tilde{S}_{\rm LLL}(\Phi,a_\mu)}/\int {\cal D}\Phi\;e^{i\tilde{S}_{\rm LLL}(\Phi,0)}\right) ,
\end{equation}
where we used ${\cal D}(u\Phi v)={\cal D}\Phi$.
Therefore, the bosonic action $\tilde{S}_{\rm LLL}(\Phi,a)$ is equivalent to the fermion action minimally coupled to abelian gauge fields
up to an irreverent normalization factor. 
By using the Polyakov-Wiegmann formula, 
\begin{equation}
\Gamma[GG^\prime]=\Gamma[G]+\Gamma[G^\prime]+\frac{1}{4\pi}\int d^2x_\parallel \; \tr\left(G^{-1}\partial_+GG^\prime\partial_-G^{\prime-1}\right),
\label{Polyakov-Wiegmann}
\end{equation}
repeatedly, we have
\begin{align}
\widetilde{S}_{\rm LLL}(\Phi,a) 
&= \Gamma[\Phi]+\frac{1}{4\pi}\int d^2x_\parallel \; \tr\left(u^{-1}\partial_+u\Phi\partial_-\Phi^{-1}+\Phi^{-1}\partial_+\Phi v\partial_-v^{-1}\right)
\notag \\
&= \frac{\NL\Nc}{8\pi}\int d^2x_{\parallel}\partial^{\mu}\Phi^{-1}\partial_{\mu}\Phi-\frac{i\NL\Nc\epsilon^{\mu\nu}}{2\pi}\int d^2x_\parallel \;e \Phi^{-1}\partial_\mu\Phi a_\nu
\label{BosonizedAction_Phi}
\end{align}
We note that for the abelian part $\Gamma[uv]$, the Wess-Zumino term is zero because the integrant is symmetric about $\mu \leftrightarrow \nu$, and $\tr$ gives the numerical factor $\NL\Nc$.
By taking $\Phi=\exp(i\sqrt{4\pi/\NL\Nc}\phi)$ and rewriting the action (\ref{BosonizedAction_Phi}) in terms of $\phi$, we obtain the abelian part of action (\ref{BosonizedAction}).
For the non-abelian part, if we define
\begin{equation}
\widetilde{S}_{\rm LLL}(h,A)\equiv \Gamma[UhV]-\Gamma[UV] ,
\end{equation}
we can show that
\begin{equation}
iW_2[A_{\mu}]=\ln\left(\int {\cal D}h\;e^{i\tilde{S}_{\rm LLL}(h,A_\mu)}/\int {\cal D}h\;e^{i\tilde{S}_{\rm LLL}(h,0)}\right) ,
\end{equation}
where we used ${\cal D}(UhV)={\cal D}h$.
By applying the Polyakov-Wiegmann formula 
to $\tilde{S}_{\rm LLL}(h,A)$ repeatedly, we obtain the non-abelian part of Eq.~(\ref{BosonizedAction}).

In terms of boson fields $h$ and $\phi$, the scalar and pseudoscalar operators, whose expectation values are the order parameters of chiral symmetry breaking, 
 are written as
\begin{align}
   \bar{c}c& =-\mu\,{\rm tr}h^{\dagger}{\rm e}^{i\sqrt{\frac{4\pi}{\NL \Nc}}\phi}-\mu\,{\rm tr}h{\rm e}^{-i\sqrt{\frac{4\pi}{\NL \Nc}}\phi}, 
   \label{scalar} \\
   \bar{c}i\gamma^5c&=-i\mu\,{\rm tr} h^{\dagger}{\rm e}^{i\sqrt{\frac{4\pi}{\NL \Nc}}\phi}
  +i\mu\,{\rm tr}h \,{\rm e}^{-i\sqrt{\frac{4\pi}{\NL \Nc}}\phi}, 
  \label{pseudoscalar}
\end{align}
 with $\mu$ being a renormalization scale~\cite{Frishman:1992mr}.
 The electric- and axial-current operators are given by
\begin{align}
  j^{\mu} &\equiv
  \bar{c}\gamma^{\mu}c=\sqrt{\frac{\NL \Nc}{\pi}}\epsilon^{\mu\nu}\partial_{\nu}\phi,
 \label{b_current}\\
 j^{\mu}_{5} &\equiv
  \bar{c}\gamma^{\mu}\gamma^5 c
  =-\sqrt{\frac{\NL \Nc}{\pi}}\partial^{\mu}\phi. 
 \label{b_a_current}
\end{align}
The corresponding operators in $(1+3)$ dimensions are related as
$\langle\bar{q}q\rangle=\langle\bar{c}c\rangle/V_\perp$, $\langle J^{\mu}\rangle=\langle j^{\mu}\rangle/V_\perp$, and so on.

 The scalar condensate at $E=0$ 
 is given by $\langle\bar{c}c\rangle_0=-2\mu|h|\cos\theta$
 with $\langle{\rm tr}h\rangle=|h|\exp(i\theta)$.
Strictly speaking, in an exactly $(1+1)$-dimensional theory, the order parameter vanishes due to the strong infrared fluctuations (the Mermin-Wagner-Coleman theorem~\cite{Mermin}).
 However, in the case of the dimensional reduction by a magnetic field, a large but finite $B$ plays a role of the infrared cutoff,
 so that the scalar condensate may stay finite~\cite{Gusynin:1995nb,Shushpanov:1997}.
 It is nontrivial whether the condensate survives at a large $B$~\cite{Fukushima:2012kc}.
 To evade this, in the following discussion, we consider only the one-flavor case.  
{There exists only one pseudoscalar meson ``$\eta$'' that would be the Nambu-Goldstone boson associated with spontaneous breaking of the U($1$) axial symmetry.
However, the U($1$) axial symmetry is explicitly broken by the axial anomaly.
As can be explicitly shown in $(1+1)$ dimensions, the pseudoscalar meson becomes massive due to the mixing with a massive photon.
 Therefore, there exists no strong infrared fluctuation of the Nambu-Goldstone boson, and thus the scalar condensate is finite.}
 In a multiflavor case, we need to assume that the scalar condensate is finite in $(1+1)$-dimensional reduced theory 
 due to some infrared cutoff.

 \section{Temporal chiral spiral}
 \label{sec:temporal chiral spiral}
 Let us consider how the scalar and pseudoscalar condensates behave 
 when an external electric field $E$ is applied  parallel to $B$. 
 The action of $\phi$ coincides with that of the Schwinger model 
 with the effective electric charge $e_{\rm eff}=e\sqrt{\NL \Nc}$~\cite{Lowenstein:1971fc}.
 First, if we keep only external electric fields and neglect the quantum fluctuation of the ``photon'' fields, we have
  \begin{align}
\langle\bar{c}c\rangle &= \langle\bar{c}c\rangle_0\cos \left(\frac{2e}{\Box}E \right), \label{cs1} \\
\langle\bar{c}i\gamma^5c\rangle &= \langle\bar{c}c\rangle_0\sin\left(\frac{2e}{\Box}E \right) \label{cs2},
 \end{align}
 where $\Box\equiv \partial_0^2-\partial_3^2$.
 If we consider a uniform electric field with a very small frequency $\Box \sim - \omega^2 \sim 0$, 
 the scalar and pseudoscalar condensates oscillate very rapidly, and their intervals become zero in the zero-frequency limit.
 This fact suggests that they have an instability against a constant electric field. 
 This instability is resolved by the screening effect from quark pairs induced by the Schwinger mechanism.

 We decompose the ``photon'' field $a_\mu$ into the quantum fluctuation $a_{q\mu}$ and the external field $a_{e\mu}$,
 \begin{equation}
  S_\phi=\int d^2x_{\parallel} \left(\frac{1}{2}\partial^{\mu}\phi\partial_{\mu}\phi 
  	- ej^{\mu} (a_{e\mu}+a_{q\mu})-\frac{V_{\perp}}{4}f^{\mu\nu}f_{\mu\nu} \right),
  \label{LLLHE3}
\end{equation}
 with $f_{\mu\nu}=\partial_{\mu}a_{q\nu}-\partial_{\nu}a_{q\mu}$.
 In $(1+1)$ dimensions, ${a}_{q\mu}$ is not a dynamical variable, and can be explicitly integrated out from the theory.
 The resultant action reads
\begin{equation}
  S_\phi=\int d^2x_{\parallel} \left(\frac{1}{2}\partial^{\mu}\phi\partial_{\mu}\phi - \frac{m_{\gamma}^2}{2}\phi^2
  	 - e j^{\mu}a_{e\mu} \right),
  \label{LLLHE4}
\end{equation}
 where $m_{\gamma}$ is the Schwinger photon mass~\cite{Lowenstein:1971fc}, 
 given by $m_{\gamma}^2\equiv e_{\rm eff}^2/(\pi V_{\perp})=e^3\Nc B/(2\pi^2)$~\cite{Kharzeev:2007jp}.
We imposed the charge neutrality condition to obtain Eq.~(\ref{LLLHE4}).
The scalar and pseudoscalar condensates become
\begin{align}
   \langle\bar{c}c\rangle & =\langle\bar{c}c\rangle_0\cos \left( \frac{2e}{\Box+m_{\gamma}^2}E \right),
  \label{cs3} \\
   \langle\bar{c}i\gamma^5c\rangle & =\langle\bar{c}c\rangle_0\sin \left( \frac{2e}{\Box+m_{\gamma}^2}E \right)
  \label{cs4}.
\end{align}
Now, the zero-frequency limit can be treated properly.

%---------------------------------------------
\begin{figure}
\includegraphics[scale=.70]{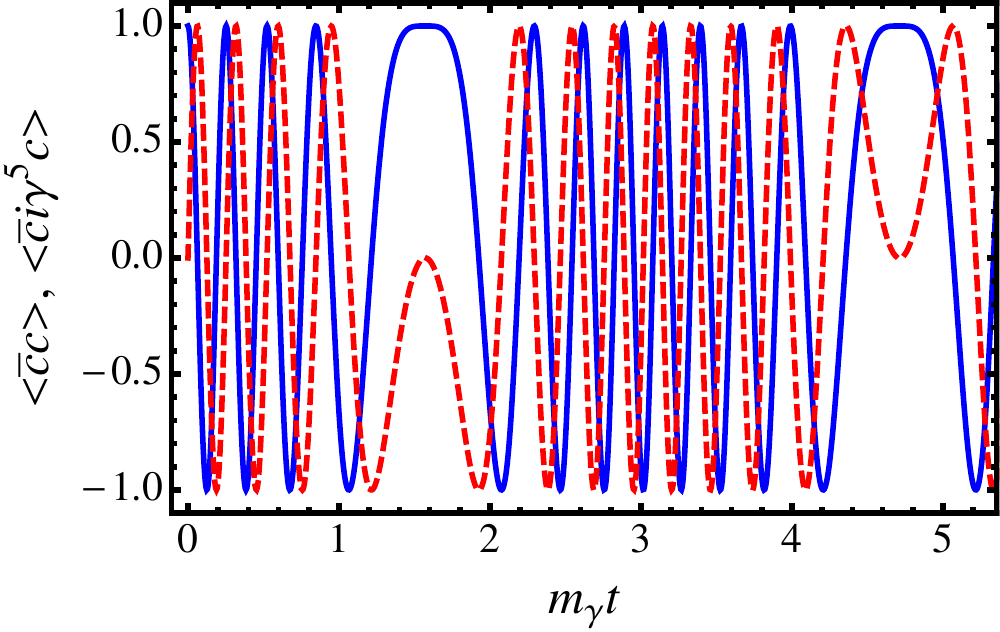}
\caption{\label{fig2}
 Scalar and pseudoscalar condensates in the instantaneous electric field $E(t)=(E_0/m_{\gamma})\delta(t)$ with $2eE_0/m_{\gamma}^2=8\pi$.  
 The blue-solid and red-dashed curves denote the scalar and pseudoscalar condensates normalized by $\langle\bar{c}c\rangle_0$, respectively.
}
\end{figure}
\begin{figure}
\includegraphics[scale=.70]{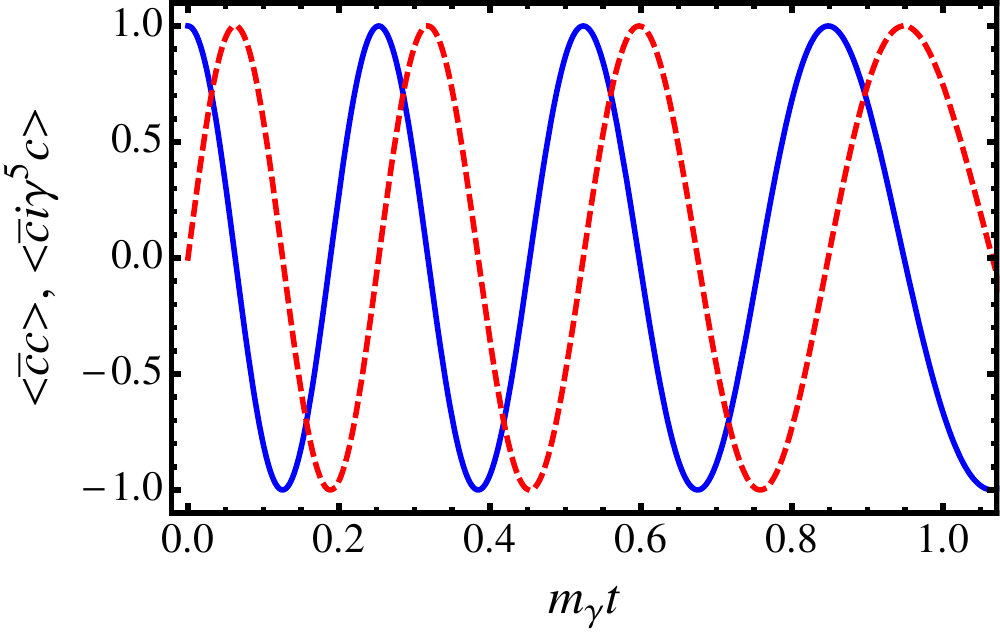}
\caption{\label{fig3}
 Scalar and pseudoscalar condensates in the vicinity of $t=0$ in the instantaneous electric field $E(t)=(E_0/m_{\gamma})\delta(t)$ with $2eE_0/m_{\gamma}^2=8\pi$. 
}
\end{figure}
%-----------------------

Here, we consider two cases of the external electric field $E(t)$: (I) the instantaneous electric field $E(t)=(E_0/m_{\gamma})\delta(t)$, and 
 (II) the constant electric field $E(t)=E_0\Theta(t)$.
For the case (I), we obtain the oscillating scalar and pseudoscalar condensates,
\begin{align}
   \langle\bar{c}c\rangle & =\langle\bar{c}c\rangle_0\cos\left(\frac{2e}{m_{\gamma}^2}E_0\sin m_{\gamma}t\right)
 \label{cs7}, \\
   \langle\bar{c}i\gamma^5c\rangle & =\langle\bar{c}c\rangle_0\sin\left(\frac{2e}{m_{\gamma}^2}E_0\sin m_{\gamma}t\right)
  \label{cs8},
\end{align}
at $t>0$, as shown in Fig.~\ref{fig2}.
 In the regions of $|m_{\gamma}t -k\pi|\ll 1$ ($k=0,1,\ldots$), they oscillate in a spiral form like 
\begin{align}
   \langle\bar{c}c\rangle & =\langle\bar{c}c\rangle_0\cos\left((-1)^k\frac{2e}{m_{\gamma}^2}E_0 \left(m_{\gamma}t-k\pi\right)\right)
 \label{cs9}, \\
   \langle\bar{c}i\gamma^5c\rangle & =\langle\bar{c}c\rangle_0\sin\left((-1)^k\frac{2e}{m_{\gamma}^2}E_0\left(m_{\gamma}t-k\pi\right)\right)
  \label{cs10}.
\end{align}
 For example, in the vicinity of $t=0$ ($k=0$), they oscillate in the standard cosine and sine waves as shown in Fig.~\ref{fig3}, 
 where the periodicity is given by $\omega=2eE_0/m_{\gamma}$.
  On the other hand,  in the regions of $|m_{\gamma}t -(k+1/2)\pi|\ll 1$, they oscillate in a spiral form, not in the standard periodic one,
 where they behave like 
\begin{align}
   \langle\bar{c}c\rangle & =\langle\bar{c}c\rangle_0\cos\left((-1)^k\frac{2e}{m_{\gamma}^2}E_0 \left(1-\left(m_{\gamma}t-\left(k+1/2\right)\pi\right)^2/2\right)\right)
 \label{cs11}, \\
   \langle\bar{c}i\gamma^5c\rangle & =\langle\bar{c}c\rangle_0\sin\left((-1)^k\frac{2e}{m_{\gamma}^2}E_0 \left(1-\left(m_{\gamma}t-\left(k+1/2\right)\pi\right)^2/2\right)\right)
  \label{cs12}.
\end{align}

\begin{figure}
\includegraphics[scale=.70]{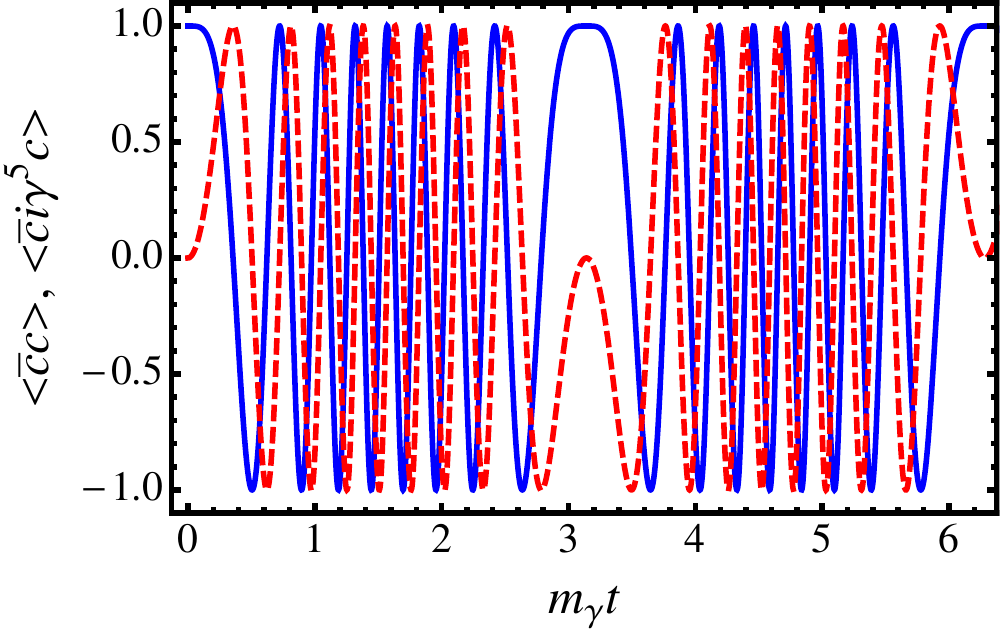}
\caption{\label{fig4}
 Scalar and pseudoscalar condensates in the constant electric field $E(t)=E_0\Theta(t)$ with $2eE_0/m_{\gamma}^2=8\pi$.  
 The blue-solid and red-dashed curves denote the scalar and pseudoscalar condensates normalized by $\langle\bar{c}c\rangle_0$, respectively.
}
\end{figure}

\begin{figure}
\includegraphics[scale=.70]{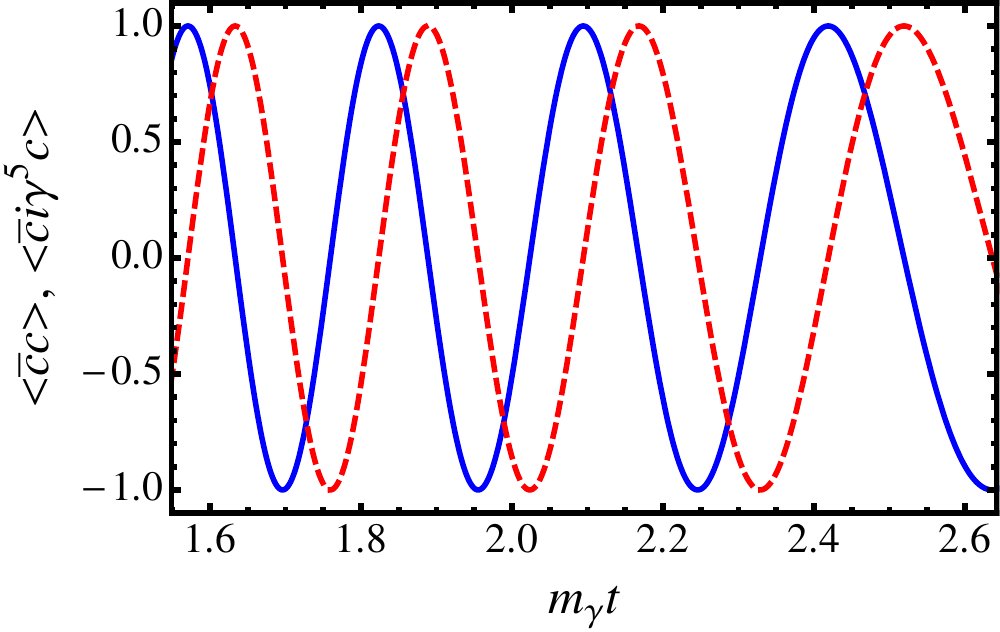}
\caption{\label{fig5}
 Scalar and pseudoscalar condensates in the vicinity of $t=\pi/2$ in the constant electric field $E(t)=E_0\Theta(t)$ with $2eE_0/m_{\gamma}^2=8\pi$. 
}
\end{figure}

In contrast,  for the case (II),
we obtain
\begin{align}
   \langle\bar{c}c\rangle & =\langle\bar{c}c\rangle_0\cos\left( \frac{2e}{m_{\gamma}^2}E_0 (1-\cos m_{\gamma}t) \right)
 \label{cs13}, \\
   \langle\bar{c}i\gamma^5c\rangle & =\langle\bar{c}c\rangle_0\sin\left( \frac{2e}{m_{\gamma}^2}E_0 (1-\cos m_{\gamma}t) \right)
  \label{cs14},
\end{align}
at $t>0$, as shown in Fig.~\ref{fig4}.
 In the regions of  $|m_{\gamma}t -k\pi|\ll 1$, they oscillate in a spiral form, not in the standard periodic one like
\begin{align}
   \langle\bar{c}c\rangle & =\langle\bar{c}c\rangle_0\cos\left(\frac{2e}{m_{\gamma}^2}E_0 \left(1+(-1)^{k+1}+(-1)^k\left(m_{\gamma}t-k\pi\right)^2/2\right)\right)
 \label{cs15}, \\
   \langle\bar{c}i\gamma^5c\rangle & =\langle\bar{c}c\rangle_0\sin\left(\frac{2e}{m_{\gamma}^2}E_0 \left(1+(-1)^{k+1}+(-1)^k\left(m_{\gamma}t-k\pi\right)^2/2\right)\right)
  \label{cs16}.
\end{align}
  On the other hand,  in the regions of $|m_{\gamma}t -(k+1/2)\pi|\ll 1$, they oscillate in a spiral form with the standard cosine and sine waves, 
  as shown in Fig.~\ref{fig5} in the case of $k=0$.
 They behave like 
\begin{align}
   \langle\bar{c}c\rangle & =\langle\bar{c}c\rangle_0\cos\left(\frac{2e}{m_{\gamma}^2}E_0 \left(1+(-1)^k\left(m_{\gamma}t-k\pi\right)\right)\right)
 \label{cs17}, \\
   \langle\bar{c}i\gamma^5c\rangle & =\langle\bar{c}c\rangle_0\sin\left(\frac{2e}{m_{\gamma}^2}E_0\left(1+(-1)^k\left(m_{\gamma}t-k\pi\right)\right)\right)
  \label{cs18},
\end{align}
 where the periodicity is given by $\omega=2eE_0/m_{\gamma}$.

%---------------------------------------------
%-----------------------

The vector and axial-vector currents are given as
  \begin{align}
     \langle j^{\mu}\rangle &= -\frac{e\NL\Nc}{\pi} \frac{\epsilon^{\mu\nu}}{\Box+m_{\gamma}^2}\partial_{\nu}E, 
    \label{cmw5} \\
      \langle j^{\mu}_5\rangle &= \frac{e\NL\Nc}{\pi} \frac{1}{\Box+m_{\gamma}^2}\partial^{\mu}E.
    \label{cmw6}
 \end{align}
 We can see the propagation of the same collective mode as the so-called ``chiral magnetic wave''~\cite{Kharzeev:2010gd}, if $m_\gamma^2$ is neglected.
From this fact, we find that in the LLL approximation, the temporal chiral spiral is nondissipative and protected by the axial anomaly.
To clarify this point, let us consider the axial anomaly in $(1+3)$ dimensions,
\begin{equation}
\partial_\mu j^\mu_{5}=\Nc\frac{eB}{2\pi}\frac{eE}{\pi} .
\label{axial anomaly}
\end{equation}
Keeping only with the contribution from the LLLs, the integrand of Eq.~(\ref{axial anomaly}) reads
\begin{equation}
\frac{d}{dt}\int dz\;j^0_{5}=\frac{e\NL\Nc}{\pi}\int dz\; E .
\label{conservation law}
\end{equation}
We note that $E$ represents the sum of external and induced electric fields in the parallel direction to $B$.
Because of a speciality of $(1+1)$ dimensions, $j^3=j_5^0$ is satisfied.
Moreover, from the integrant of the Maxwell equation in $(1+1)$ dimensions, $V_{\perp}\partial_{0}f^{03}=ej^3$, we have
\begin{equation}
\frac{d}{dt}\int dz\; E=-\frac{e}{V_{\perp}}\int dz\;j^3.
\label{Maxwell equation}
\end{equation}
From Eqs.~(\ref{conservation law}) and~(\ref{Maxwell equation}), 
the electric current and total electric field become harmonic oscillators with the same frequency $m_{\gamma}$, which are not attenuated permanently~\cite{Iwazaki}.
The oscillation of the electric current can also be understood as the propagation of the chiral magnetic wave as shown in Eqs.~(\ref{cmw5}) and~(\ref{cmw6}), 
and thus the chiral magnetic wave must be nondissipative.
Then, since the temporal oscillation of scalar and pseudoscalar condensates also originates with the propagation of the chiral magnetic wave as shown in Eqs.~(\ref{cs3}) and ~(\ref{cs4}), the temporal chiral spiral is also nondissipative. 
We emphasize that this nondissipative nature comes from the continuity equation of axial charge in Eq.~(\ref{conservation law}), or equivalently, the axial anomaly keeping only with the LLL contributions. 
Therefore, the temporal chiral spiral may be nondissipative even at finite temperature 
as long as the LLL approximation is plausible, because the axial anomaly is not modified by temperatures~\cite{Fukushima:2012vr}.
If the corrections from higher Landau levels are taken into account, 
the chiral magnetic wave may become dissipative~\cite{Kharzeev:2010gd,Metlitski:2005pr}.
Then, the scalar and pseudoscalar condensates behave as damping oscillators with a spiral structure. 

Finally, let us discuss the temporal chiral spiral in a dense quark matter.
 To make a quark number density finite, we introduce the external current to the action (\ref{LLLaction}).
 The action reads
 \begin{equation}
  S_{\mathrm{LLL}}= \int d^2x_{\parallel}\;\Bigl(i\bar{c}_f\gamma^{\mu}(\partial_{\mu}+iea_{\mu})c_f+ej_{\rm ext}^{\mu}a_\mu
    +\bar{c}_f\gamma^{\mu}[A_{\mu}]_{ff^{\prime}}c_{f^{\prime}}\Bigr) .
 \label{LLLaction2}
\end{equation}
When the external current is $j_{\rm ext}^\mu=(\NL\Nc\rho,0)$, the boson field $\phi$ is shifted as $\phi=\phi^{\prime}+\sqrt{\pi\NL\Nc}\rho z$.
The dynamics of the new field $\phi^{\prime}$ is the same as before, and the effect of $\rho$ on the expectation values of quark composite operators
 comes only through the local shift of $\phi$ in Eqs.~(\ref{scalar}),~(\ref{pseudoscalar}),~(\ref{b_current}) and~(\ref{b_a_current}).
For example, when the instantaneous external electric field is applied,
the real-time evolution of the scalar and pseudoscalar condensates are
\begin{align}
      \langle\bar{c}c\rangle &=\langle\bar{c}c\rangle_0\cos\left(\frac{2e}{m_{\gamma}^2}E_0\sin m_{\gamma}t+2\pi\rho z\right), \\
      \langle\bar{c}i\gamma^5c\rangle&=\langle\bar{c}c\rangle_0\sin\left(\frac{2e}{m_{\gamma}^2}E_0\sin m_{\gamma}t+2\pi\rho z\right),
 \end{align}
which exhibit the coexistence of the temporal and spatial chiral spirals.
The quark number density is $\langle j^0 \rangle = \NL\Nc\rho$, which is independent of $t$ and $z$.
The temporal and spatial chiral spirals form a propagating wave.
In the regions of $|m_{\gamma}t -k\pi|\ll 1$, 
the dispersion relation is linear, whose velocity is given as $v=eE_0/(\pi m_{\gamma}\rho)$.
 In a charge neutral matter, such as in a neutron star, we need to take into account background charges, such as electrons or muons.
 Assuming that the dynamical scales of quarks and the background charges are sufficiently separated and the background charges are approximated to be static, we can apply the same argument~\cite{Alford:2002kj}.

 \section{Concluding remarks}
 \label{sec:concluding remarks}
 In this paper, we have presented the analysis of dynamical properties of the scalar and pseudoscalar condensates
  in finite electric and strong magnetic fields. 
 In particular, we have discussed a temporal analog to the spatial chiral spiral in the LLL approximation.
 Although the scalar and pseudoscalar condensates seem unstable against a constant external electric field, 
 this instability is resolved by the backreaction from quark pairs induced by the Schwinger mechanism.
 We showed that the uniform electric field that is
 instantaneously applied parallel to a strong magnetic field
 induces a temporal oscillation of the scalar and pseudoscalar condensates in a spiral form.

 There are several future directions toward QCD phenomenology.
 In the quark-gluon plasma with a strong magnetic field or
 the inner core of magnetars, the temporal chiral spiral plays a key role for the dynamical restoration of chiral symmetry.
 In realistic situations where magnetic fields do not reach the scale that the LLL approximation is plausible, 
 the temporal chiral spiral may be damped by dissipative modes.
 Therefore, we need to treat the corrections beyond the LLL and check whether the temporal chiral spiral survives.
 We can also investigate in detail the temporal chiral spiral at finite temperatures and densities
 on the basis of effective models of QCD. 
 Another important subject is a time-periodic charged meson condensate in two-flavor QCD.

 Our analysis can be applied to quasi-one-dimensional condensed matter systems
 described by the Tomonaga-Luttinger liquid~\cite{Tomonaga:1950}.
 The time-periodic condensate to be addressed in this paper
 would also be realized in, e.g., quantum wires like carbon nanotubes~\cite{Bockrath:1999},
 fractional quantum Hall edge states~\cite{Wen:1990}, and
 one-dimensional cold-atom systems~\cite{Guan:2013}.
 For example, in the case of carbon nanotubes,
 the Schwinger photon mass would be scaled by the Fermi velocity $v_F$ as $m_{\gamma}^2=g^2e^2/(v_F\pi^2 R^2)$ with $g$
 being a parameter of Tomonaga-Luttinger liquid and $R$ being a tube radius~\cite{Kane:1992}, respectively.
 We can estimate the magnitude of oscillations, 
 $\omega=2eE_0/m_{\gamma}\sim 10^{10}$ Hz with $v_{F}=8\times10^5$ m/s, $g=0.2$, $R=1$ nm~\cite{Kane:1992,Ishii:2003},
 and $eE_0=1 \, {\rm meV}/\mu{\rm m}$.
 In these condensed matter systems, one can discuss the experimental observation of the ``time crystal,''
 namely, the ground state with time-periodic condensates~\cite{Wilczek:2012jt},
 by utilizing the nondissipative collective mode protected by quantum anomaly,
 instead of spontaneous symmetry breaking.
 
\begin{acknowledgements}

 We thank T.~Kojo, Y.~Tanizaki, and H.~Watanabe for stimulating discussions and helpful comments.
 T.~H.~was supported by JSPS Research Fellowships for Young Scientists.
A.~Y.~was supported by the Special Postdoctoral Research Program of RIKEN.
 This work was partially supported by JSPS KAKENHI Grants No. 24740184 and No. 23340067.
This work was also partially supported by RIKEN iTHES Project.

\end{acknowledgements}

\appendix{}

\section{Landau quantization of relativistic particles}
\label{sec: propagator}

\subsection{Scalar field}
We consider the equation of motion for the charged scalar field $\Phi$ in a uniform magnetic field,
\begin{equation}
(D^{\mu}D_{\mu}+m^2)\Phi(x)=0 ,
\label{eom1: charged scalar}
\end{equation}
with the background covariant derivative
$D_{\mu}=\partial_{\mu}+ieA^{\rm em}_{\mu}$.
Taking the symmetric gauge $A^{\rm em}_{\mu}=(0,By/2,-Bx/2,0)$, we rewrite Eq.~(\ref{eom1: charged scalar}) with an ansatz $\Phi(x)=f(x_{\parallel})\varphi(x_{\perp})$ as
\begin{align}
 (\partial_0^2-\partial_z^2+m^2)f(x_{\parallel})&=-\lambda f(x_{\parallel}), \label{eom2: charged scalar}\\
 -( (D_1)^2+ (D_2)^2 )\varphi(x_{\perp})&=\lambda \varphi(x_{\perp}) .
\label{eom3: charged scalar}
\end{align}
From Eq.~(\ref{eom2: charged scalar}), we can write $f(x_{\parallel})=\exp(-i\varepsilon t+ip^3z)$ and $\lambda=\varepsilon^2-(p^3)^2-m^2$.
We can easily solve  Eq.~(\ref{eom3: charged scalar}) by rewriting it in the Dirac bracket form,
\begin{equation}
eBH|\varphi\rangle=\lambda |\varphi\rangle ,
\label{eom4: charged scalar}
\end{equation}
where $H\equiv-( (D_1)^2+(D_2)^2 )/eB$. 
We introduce new position and momentum operators as
\begin{align}
  & X\equiv\frac{1}{\sqrt{eB}}(i\partial_y+xeB/2), \quad P_X\equiv\frac{1}{\sqrt{eB}}(-i\partial_x+yeB/2) , \\
  & Y\equiv\frac{1}{\sqrt{eB}}(i\partial_x+yeB/2), \quad  P_Y\equiv\frac{1}{\sqrt{eB}}(-i\partial_y+xeB/2) .
\end{align}
They satisfy the canonical commutation relations $[X,P_X]=i$, $[Y,P_Y]=i$, and others become zero.
Then, $H$ reads
\begin{equation}
H=P_X^2+X^2 .
\label{eom5: charged scalar}
\end{equation}
The transverse motion of the charged scalar field becomes a harmonic oscillator and thus quantized, which is the so-called Landau quantization.
Now, we introduce creation and annihilation operators
\begin{align}
  & a\equiv\frac{1}{\sqrt{2}}(X+iP_X),  \quad a^{\dagger}\equiv\frac{1}{\sqrt{2}}(X-iP_X) , \\
  & b\equiv\frac{1}{\sqrt{2}}(Y+iP_Y), \quad  b^{\dagger}\equiv\frac{1}{\sqrt{2}}(Y-iP_Y) ,
  \label{eq:creationAnihilationOperators}
\end{align}
where $[a,a^\dag]=[b,b^\dag]=1$, and other commutation relations vanish.
Using these operators, the Hamiltonian is written as
\begin{equation}
H=2a^{\dagger}a+1.
\label{eom6: charged scalar}
\end{equation}
Since the eigenvalue of $H$ is $2n+1$ with integer $n$, the energy of particles and antiparticles is given by
\begin{equation}
\varepsilon(n,p^3) =\pm\sqrt{(p^3)^2+m^2+eB(2n+1)} .
\label{landau level1}
\end{equation}
In the symmetric gauge, the angular momentum $L_z$ is given by 
\begin{equation}
L_z=b^{\dagger}b-a^{\dagger}a ,
\end{equation}
and it apparently commutes with $H$.
Therefore, the eigenstates of Eq.~(\ref{eom4: charged scalar}) are labeled by the Landau level $n$ and the angular momentum $l$ as
\begin{equation}
|n,l\rangle=\frac{1}{\sqrt{n!(n+l)!}}(a^{\dagger})^n(b^{\dagger})^{n+l}|0,0\rangle,
\end{equation}
where $|0,0\rangle$ is defined by $a|0,0\rangle=b|0,0\rangle=0$.
We note that $l$ has the minimum $l=-n$, but it is not bounded from above by $n$. 

Let us now derive the wave function of $|n,l\rangle$ in the coordinate space.
We define the complex coordinates
\begin{equation}
 \cz\equiv\sqrt{\frac{eB}{2}}(x+iy), \quad  \bar{\cz}\equiv\sqrt{\frac{eB}{2}}(x-iy) .
 \label{eq:complexVariables}
\end{equation}
From Eqs.~(\ref{eq:creationAnihilationOperators}) and (\ref{eq:complexVariables}),
the creation and annihilation operators can be written as 
\begin{align}
a&= \frac{\cz}{2}+\partial_{\bar{\cz}}, \quad a^\dag=\frac{\bar{\cz}}{2}-\partial_\cz,\\
b&=\frac{\bar{\cz}}{2}+\partial_\cz,\quad
b^\dag=\frac{\cz}{2}-\partial_{\bar{\cz}}.
\end{align}
The LLL wave function with $l=0$ is obtained by $a|0,0\rangle=b|0,0\rangle=0$, which are 
\begin{equation}
\left(\frac{\cz}{2}+\partial_{\bar{\cz}}\right) \varphi_{0,0}(\cz,\bar{\cz}) = \left(\frac{\bar{\cz}}{2}+\partial_\cz\right) \varphi_{0,0}(\cz,\bar{\cz})=0
\end{equation}
in the coordinate space. Here $\varphi_{0,0}(\cz,\bar{\cz}) \equiv \langle \cz,\bar{\cz}|0,0\rangle$.
The solution is
\begin{equation}
\begin{split}
\varphi_{0,0}(\cz,\bar{\cz})  = \sqrt{\frac{eB}{2\pi}}e^{-\frac{|\cz|^2}{2}},
\end{split}
\end{equation}
where the normalization of the wave function is chosen as
\begin{equation}
\int \frac{d\cz d\bar{\cz}}{eB}\varphi^*_{n,l}(\cz,\bar{\cz})\varphi_{n,l}(\cz,\bar{\cz})=1.
\end{equation}
The wave function of $|n,l\rangle$ in the complex coordinates reads
\begin{align}
\varphi_{n,l}(\cz,\bar{\cz})\equiv\langle \cz,\bar{\cz}|n,l\rangle &= \frac{1}{\sqrt{n!(n+l)!}}\langle \cz,\bar{\cz}|(a^{\dagger})^n(b^{\dagger})^{n+l}|0,0\rangle  
\notag \\
&= \frac{1}{\sqrt{n!(n+l)!}}\left(\frac{\bar{\cz}}{2}-\partial_{\cz}\right)^n\left(\frac{\cz}{2}-\partial_{\bar{\cz}}\right)^{n+l}\varphi_{0,0}(\cz,\bar{\cz})
\notag \\
&= \frac{1}{\sqrt{n!(n+l)!}}e^{\frac{|\cz|^2}{2}}\left(-\partial_{\cz}\right)^n\left(-\partial_{\bar{\cz}}\right)^{n+l}e^{-\frac{|\cz|^2}{2}}\varphi_{0,0}(\cz,\bar{\cz}) 
\notag \\
&= \frac{1}{\sqrt{n!(n+l)!}}e^{\frac{|\cz|^2}{2}}(-1)^n\frac{1}{\bar{\cz}^l}\left(\frac{\partial}{\partial |\cz|^2}\right)^n|\cz|^{2(n+l)}e^{-\frac{|\cz|^2}{2}}\varphi_{0,0}(\cz,\bar{\cz}) 
\notag \\
&= \sqrt{\frac{eB}{2\pi}}\sqrt{\frac{n!}{(n+l)!}}(-1)^n\cz^le^{-\frac{|\cz|^2}{2}}L_n^{(l)}(|\cz|^2) ,
\label{wave function: scalar}
\end{align}
where we used
the generalized Laguerre polynomials
\begin{equation}
 L_n^{(l)}(x)=\frac{e^xx^{-l}}{n!}\frac{d^n}{dx^n}x^{n+l}e^{-x}.
\end{equation}
Taking $n=0$, we obtain the LLL wave function
 \begin{equation}
\varphi_l(x_\perp)=\sqrt{\frac{eB}{2\pi l!}}\left(\frac{eB}{2}\right)^{\frac{l}{2}}(x+iy)^le^{-\frac{eB}{4}(x^2+y^2)}.
\label{LLL wave function}
\end{equation} 

\subsection{Dirac field}
Next, we consider a Dirac field $\Psi$ in a uniform magnetic field.
The Dirac equation in $(1+3)$ dimensions is
\begin{equation}
(i\gamma^{\mu}D_{\mu}-m)\Psi(x)=0.
\label{eom1: Dirac}
\end{equation}
The (1+3)-dimensional gamma matrices in the Dirac representation are 
\begin{equation}
 \gamma^0=
 \begin{pmatrix}
  1 & 0 \\
  0 & -1
 \end{pmatrix} , \quad
\gamma^i=
 \begin{pmatrix}
  0 & \sigma^i \\
  -\sigma^i & 0
 \end{pmatrix} .
\end{equation}
Equation~(\ref{eom1: Dirac}) in the Dirac representation is explicitly  
\begin{equation}
\begin{pmatrix}
  iD_0-m & i\sigma^iD_i \\
  -i\sigma^iD_i & -iD_0-m
 \end{pmatrix}
\begin{pmatrix}
  u \\
  \bar{v}
 \end{pmatrix}
=0,
\label{eom2: Dirac}
\end{equation}
where $\sigma^i$ are the Pauli matrices.
When we choose the symmetric gauge, $u$ satisfies
\begin{align}
(D_0^2+m^2)u &= \left(\sigma^iD_i\right)^2u
\notag \\
&= -\left(D^iD_i+eB_i\sigma^i\right)u ,
\label{eom3: Dirac}
\end{align}
where we used $\sigma^i\sigma^j=i{\epsilon^{ij}}_k\sigma^k+\delta^{ij}$, and $B_i=-{\epsilon_i}^{jk}F_{jk}/2$.
Taking an ansatz $u(x)=\psi_{\pm}(x)\chi_{\pm}$ with $\sigma^3 \chi_{\pm}=\pm\chi_{\pm}$, we have
\begin{equation}
\left(D^{\mu}D_{\mu}+m^2\mp eB\right)\psi_{\pm}(x) = 0 .
\label{eom4: Dirac}
\end{equation}
The energy of particles and antiparticles (with spin $s=\pm1/2$) is discretized as
\begin{equation}
\varepsilon(n,p^3,s)=\pm\sqrt{{(p^3)}^2+m^2+2eB(n+1/2\mp s)}.
\end{equation}
The wave functions are labeled by the energy level $n$, the angular momentum $l$, the momentum parallel to the magnetic field, $p^3$, and the spin $s=\pm1/2$ as
\begin{align}
|\Psi(n,l,p^3,s)\rangle &=
 \begin{pmatrix}
  \sqrt{\varepsilon+m}\chi_{s}|n,l\rangle|p^3\rangle\\
  \frac{-i\sigma^i D_i}{\sqrt{\varepsilon+m}}\chi_{s}|n,l\rangle|p^3\rangle
  \end{pmatrix} 
  \notag \\
&=
 \begin{pmatrix}
  \sqrt{\varepsilon+m} \chi_{s} |n,l\rangle|p^3\rangle \\
  \frac{1}{\sqrt{\varepsilon+m}}\left(2p^3s+i\sqrt{2eB}(\sigma^+a^{\dagger}-\sigma^-a)\right) \chi_{s} |n,l\rangle|p^3\rangle
  \end{pmatrix}   ,
\label{wave function: Dirac}
\end{align}
with $\sigma^{\pm}$ being defined as
\begin{equation}
 \sigma^+=\frac{1}{2}(\sigma^1+i\sigma^2)= 
 \begin{pmatrix}
  0 & 1 \\
  0 & 0
 \end{pmatrix},\quad 
 \sigma^-=\frac{1}{2}(\sigma^1-i\sigma^2)=
 \begin{pmatrix}
  0 & 0 \\
  1 & 0
 \end{pmatrix} .
\end{equation}
$|p^3\rangle$ is the eigenstate of $i\partial_z$ with eigenvalue $p^3$.
For given $\bar{\varepsilon}\equiv\sqrt{(p^3)^2+m^2+2eBn}$, the wave functions in Eq.~(\ref{wave function: Dirac}) with the spin $\pm1/2$ are explicitly
\begin{equation}
|\Psi(n,l,p^3,+)\rangle =
 \begin{pmatrix}
  \sqrt{\bar{\varepsilon}+m}|n,l\rangle|p^3\rangle\\
  0 \\
  \frac{p^3}{\sqrt{\bar{\varepsilon}+m}}|n,l\rangle |p^3\rangle\\
  -i\sqrt{\frac{2eB}{\bar{\varepsilon}+m}}a|n,l\rangle|p^3\rangle
  \end{pmatrix} ,
\quad
|\Psi(n,l,p^3,-)\rangle =
 \begin{pmatrix}
  0 \\
  \sqrt{\bar{\varepsilon}+m}|n-1,l\rangle|p^3\rangle\\
  i\sqrt{\frac{2eB}{\bar{\varepsilon}+m}}a^{\dagger}|n-1,l\rangle|p^3\rangle \\
  -\frac{p^3}{\sqrt{\bar{\varepsilon}+m}}|n-1,l\rangle |p^3\rangle
  \end{pmatrix} .
\end{equation}
In the spin basis, the wave functions are written as
\begin{align}
|\Psi'(n,l,p^3,+)\rangle &\equiv U|\Psi(n,l,p^3,+)\rangle=
 \begin{pmatrix}
  \sqrt{\bar{\varepsilon}+m}|n,l\rangle|p^3\rangle\\
  \frac{p^3}{\sqrt{\bar{\varepsilon}+m}}|n,l\rangle|p^3\rangle \\
  0 \\
  -i\sqrt{\frac{2eB}{\bar{\varepsilon}+m}}a|n,l\rangle|p^3\rangle
  \end{pmatrix} ,\\
|\Psi'(n,l,p^3,-)\rangle& \equiv U|\Psi(n,l,p^3,-)\rangle=
 \begin{pmatrix}
  0 \\
  i\sqrt{\frac{2eB}{\bar{\varepsilon}+m}}a^{\dagger}|n-1,l\rangle |p^3\rangle\\
  \sqrt{\bar{\varepsilon}+m}|n-1,l\rangle|p^3\rangle\\
  -\frac{p^3}{\sqrt{\bar{\varepsilon}+m}}|n-1,l\rangle |p^3\rangle
  \end{pmatrix} .
\end{align}
where $U$ is a unitary matrix,
\begin{equation}
U= 
\begin{pmatrix}
1&0&0&0\\
0&0&1&0\\
0&1&0&0\\
0&0&0&1
\end{pmatrix}.
\end{equation}
In this basis, the wave function of the LLL becomes
\begin{equation}
|\Psi'(0,l,p^3,+)\rangle =
 \begin{pmatrix}
  \sqrt{\bar{\varepsilon}+m}|0,l\rangle|p^3\rangle\\
  \frac{p^3}{\sqrt{\bar{\varepsilon}+m}}|0,l\rangle|p^3\rangle \\
  0 \\
0
  \end{pmatrix} .
\end{equation}

Here, let us derive the propagator. The sum of the spin and angular momentum is
\begin{align}
&\sum_{l,s}|\Psi(n,l,p^3,s)\rangle\langle\Psi(n,l,p^3,s)|\\
 &\quad= \sum_{l=-n}^\infty|\Psi(n,l,p^3,+)\rangle\langle\Psi(n,l,p^3,+)|+\sum_{l=-n+1}^{\infty}|\Psi(n,l,p^3,-)\rangle\langle\Psi(n,l,p^3,-)|
\notag \\
&\quad= (\gamma^0\bar{\varepsilon}+\gamma^3p_3+m)|p^3\rangle\langle p^3|\left(P_+\sum_{l=-n}^\infty|n,l\rangle\langle n,l|+P_-\sum_{l=-n+1}^\infty|n-1,l\rangle\langle n-1,l|\right) 
\notag \\
&\qquad  -\frac{i}{2}\sqrt{2eB}\left(\gamma^1+i\gamma^2\right)|p^3\rangle\langle p^3|\sum_{l=-n}^\infty|n,l\rangle\langle n,l|a^{\dagger}\notag\\
&\qquad+\frac{i}{2}\sqrt{2eB}\left(\gamma^1-i\gamma^2\right)|p^3\rangle\langle p^3|\sum_{l=-n}^\infty a|n,l\rangle\langle n,l| ,
\end{align}
where we used
\begin{align}
\sum_{l=-n}^\infty\;a|n,l\rangle\langle n,l|a^\dagger&=\sum_{l=-n+1}^{\infty}\;n|n-1,l\rangle\langle n-1,l| ,\\
\sum_{l=-n+1}^\infty\;a^\dagger|n-1,l,\rangle\langle n-1,l|a&=\sum_{l=-n}^{\infty}\;n|n,l\rangle\langle n,l| ,\\
\sum_{l=-n}^\infty\;a|n,l\rangle\langle n,l|&=\sum_{l=-n+1}^{\infty}\;|n-1,l\rangle\langle n-1,l|a ,\\
\sum_{l=-n}^\infty\;|n,l\rangle\langle n,l|a^\dagger&=\sum_{l=-n+1}^{\infty}\;a^\dagger|n-1,l\rangle\langle n-1,l| .
\end{align}
$P_{\pm}=(1\pm i\gamma^1\gamma^2)/2$ is the projection operator of the spin along  the magnetic axis.
The fermion propagator is obtained from the spectral decomposition
\begin{equation}
S(x_1,x_2)=\sum_{n,l,s}\int\frac{dp^0}{2\pi}e^{-ip^0(t_1-t_2)}\int\frac{dp^3}{2\pi}\frac{\langle x_1|\Psi(n,l,p^3,s)\rangle \langle\Psi(n,l,p^3,s)|x_2\rangle}{(p^0)^2-\bar{\varepsilon}^2} .
\label{eq:propagator1}
\end{equation}
To perform the sum of the angular momentum and spin in the denominator of Eq.~(\ref{eq:propagator1}), we need the following relations:
\begin{align}
\sum_{l=-n}^\infty\;\langle \cz_1,\bar{\cz}_1|n,l\rangle\langle n,l|\cz_2,\bar{\cz}_2\rangle &= \frac{eB}{2\pi}U(\cz_1,\cz_2)e^{-\frac{|\cz_1-\cz_2|^2}{2}}L^{(0)}_n(|\cz_1-\cz_2|^2) ,\\
\sum_{l=-n}^\infty\;\langle \cz_1,\bar{\cz}_1|a|n,l\rangle\langle n,l|\cz_2,\bar{\cz}_2\rangle 
&= \left(\frac{\cz_1}{2}+\partial_{\bar{\cz}_1}\right)\frac{eB}{2\pi}U(\cz_1,\cz_2)e^{-\frac{|\cz_1-\cz_2|^2}{2}}L^{(0)}_n(|\cz_1-\cz_2|^2) 
\notag \\
&= \frac{eB}{2\pi}U(\cz_1,\bar{\cz}_1,\cz_2,\bar{\cz}_2)\left(\cz_2-\cz_1\right)e^{-\frac{|\cz_1-\cz_2|^2}{2}}L^{(1)}_{n-1}(|\cz_1-\cz_2|^2) , \\
\sum_{l=-n}^\infty\;\langle \cz_1,\bar{\cz}_1|n,l\rangle\langle n,l|a^{\dagger}|\cz_2,\bar{\cz}_2\rangle 
&= \frac{eB}{2\pi}U(\cz_1,\bar{\cz}_1,\cz_2,\bar{\cz}_2)\left(\bar{\cz}_1-\bar{\cz}_2\right)e^{-\frac{|\cz_1-\cz_2|^2}{2}}L^{(1)}_{n-1}(|\cz_1-\cz_2|^2) ,
\end{align}
where $U(\cz_1,\bar{\cz}_1,\cz_2,\bar{\cz}_2)$ is a nonlocal phase factor depending on the choice of gauge given by
\begin{equation}
U(\cz_1,\bar{\cz}_1,\cz_2,\bar{\cz}_2)=e^{\frac{1}{2}(\cz_1\bar{\cz}_2-\bar{\cz}_1\cz_2)} .
\end{equation}
The propagator can be decomposed into $U(\cz_1,\bar{\cz}_1,\cz_2,\bar{\cz}_2)$ and a function depending on $x_1-x_2$,
\begin{equation}
\begin{split}
S(x_1,x_2)= U(\cz_1,\bar{\cz}_1,\cz_2,\bar{\cz}_2) \tilde{S}(x_1-x_2).
\end{split}
\end{equation}
In momentum space, $\tilde{S}$ can be written as~\cite{propagator}

\begin{equation}
  \tilde{S}(p)=e^{-\frac{(p_\perp)^2}{eB}}\sum_{n=0}^{\infty}\;2(-1)^n\frac{D_{n}(p)}{{p_0}^2-\bar{\varepsilon}^2}, 
\label{propagator1}
\end{equation}
 with
\begin{align}
  D_{n}(p)&=(\gamma^0p_0+\gamma^3p_3+m)\left(P_+L^{(0)}_n\left( \frac{2(p_{\perp})^2}{eB} \right)-P_-L^{(0)}_{n-1}\left( \frac{2(p_{\perp})^2}{eB} \right)\right)\notag\\
  &\quad+2(\gamma^1p_1+\gamma^2p_2)L^{(1)}_{n-1}\left( \frac{2(p_{\perp})^2}{eB} \right).
\label{propagator2}
\end{align}
 The important fact is that the transverse dynamics of all the Landau levels is suppressed if $(p_{\perp})^2 \equiv (p^1)^2+(p^2)^2 \ll eB$, because the transverse momentum is scaled as $(p_{\perp})^2/eB$ in Eqs.~(\ref{propagator1}) and (\ref{propagator2}).

\section{Generating functional in $(1+1)$ dimensions}
\label{sec: nonabelian bosonization}

First, we calculate the determinant of the Dirac operator in the two-dimensional Euclidean space $x_\parallel = (x_3,x_4)$, where $x_3$ is a spatial coordinate parallel to the magnetic field and $x_4$ is an imaginary time.
We use the convention of 
\begin{align}
\gamma_3\equiv
 \begin{pmatrix}
  0 & i \\
  -i & 0
 \end{pmatrix} &, \quad
\gamma_4\equiv
 \begin{pmatrix}
  0 & 1 \\
  1 & 0
 \end{pmatrix} , 
 \notag \\
 \gamma_+\equiv\gamma_3+i\gamma_4=
 \begin{pmatrix}
  0 & 2i \\
  0 & 0
 \end{pmatrix} &, \quad
\gamma_-\equiv\gamma_3-i\gamma_4=
 \begin{pmatrix}
  0 & 0 \\
  -2i & 0
 \end{pmatrix},
\end{align}
and $\gamma_5=-i\gamma_3\gamma_4$.
They satisfy 
\begin{align}
\{\gamma_\mu,\gamma_\nu\}&=2\delta_{\mu\nu}, \\
[\gamma_\mu,\gamma_\nu]&=2i\epsilon_{\mu\nu}\gamma_5 ,
\end{align}
where $\epsilon_{\mu\nu}$ are totally antisymmetric tensors in the Euclidean space with $\epsilon_{34}=1$.
The Euclidean action is
\begin{equation}
  S_{\mathrm{E}}= \int d^2x_{\parallel}\;\bar{c} \SlD c.
 \label{Euclidean action}
\end{equation}
with the Dirac operator
\begin{equation}
\SlD=\gamma_{\mu}\left(\partial_{\mu}+iea_{\mu}-iA_{\mu}\right) .
\end{equation}
The generating functional $W_{\rm E}[a_{\mu},A_{\mu}]$ is defined as
\begin{equation}
  W_{\rm E}[a_{\mu},A_{\mu}]\equiv   -\ln \int{\cal D}\bar{c}{\cal D}c\;e^{-S_{\mathrm{E}}}
  =-\ln{\rm det}(-\Sl{D}) .
 \label{Euclidean generating functional}
\end{equation}
In an infinite volume, the gauge fields can be decomposed as
\begin{align}
-ea_{\mu}&=\frac{i}{2}(\delta_{\mu\nu}-i\epsilon_{\mu\nu})v\partial_{\nu}v^{-1}+\frac{i}{2}(\delta_{\mu\nu}+i\epsilon_{\mu\nu})u^{-1}\partial_{\nu}u,
\label{U(1)_E}\\
A_{\mu}&=\frac{i}{2}(\delta_{\mu\nu}-i\epsilon_{\mu\nu})V\partial_{\nu}V^{-1}+\frac{i}{2}(\delta_{\mu\nu}+i\epsilon_{\mu\nu})U^{-1}\partial_{\nu}U .
\label{SU(NLNc)_E}
\end{align}
From the hermitcity of gauge fields,  
\begin{align}
u^{\dagger}=v & , \quad U^{\dagger}=V
\label{hermiticity}
\end{align}
are satisfied.
We have
\begin{align}
 -e\gamma_{\mu}a_{\mu}&= \frac{i}{2}\left(\gamma_{+}v\partial_{-}v^{-1}+\gamma_{-}u^{-1}\partial_{+}u \right),\\
 \gamma_{\mu}A_{\mu}&= \frac{i}{2}\left(\gamma_{+}V\partial_{-}V^{-1}+\gamma_{-}U^{-1}\partial_{+}U \right),
\end{align}
where $\partial_{\pm}\equiv\partial/\partial x_3 \pm i\partial/\partial x_4$. 
The Dirac operator is written as 
\begin{equation}
\SlD=\gamma_{\mu}(\partial_{\mu}+iea_{\mu}-iA_{\mu})=\Omega \gamma_{\mu}\partial_{\mu}\overline{\Omega} ,
\label{Dirac operator}
\end{equation}
where $\Omega$ and $\overline{\Omega}$ are given by
\begin{equation}
 \Omega=
 \begin{pmatrix}
  vV & 0 \\
  0 & u^{-1}U^{-1}
 \end{pmatrix} , \quad
\overline{\Omega}=
 \begin{pmatrix}
  uU & 0 \\
  0 & v^{-1}V^{-1}
 \end{pmatrix}
.
\label{Omega}
\end{equation}

Let us define a Dirac operator $\SlDr$ parametrized by one real parameter $r\in[0,1]$ as
\begin{equation}
\SlDr=\gamma_{\mu}\left(\partial_{\mu}+iea_{\mu}^{(r)}-iA_{\mu}^{(r)}\right)=\Omega^{(r)} \gamma_{\mu}\partial_{\mu}\overline{\Omega}^{(r)} ,
\label{Dirac_r}
\end{equation}
where $a_{\mu}^{(1)}=a_{\mu}$, $A_{\mu}^{(1)}=A_{\mu}$, and $a_{\mu}^{(0)}=A_{\mu}^{(0)}=0$.
We also define
\begin{equation}
\omega(r)\equiv \ln{\rm det}\left(\left[\SlDr\right]^{\dagger}\SlDr \right).
\end{equation}
In the proper-time regularization, $\omega(r)$ reads
\begin{equation}
\omega(r)=-\int_\delta^\infty\frac{d \tau}{\tau}\mathop{\mathrm{Tr}} e^{-\tau\left[\SlDr\right]^{\dagger}\SlDr} ,
\end{equation}
where $\delta$ is the regularization parameter.
The capital trace ``Tr'' runs over the coordinate, color, and spinor spaces.
The derivative of $\omega(r)$ with respect to $r$ is given by
\begin{align}
\frac{d\omega(r)}{d r}&= \int_\delta^\infty d \tau\mathop{\mathrm{Tr}} \left(\frac{d\left[\SlDr\right]^{\dagger}}{d r}\SlDr+\left[\SlDr\right]^{\dagger}\frac{d\SlDr}{d r}\right)e^{-\tau\left[\SlDr\right]^{\dagger}\SlDr} 
\notag \\
&= \int_\delta^\infty d \tau\mathop{\mathrm{Tr}}\Bigl(W^{(r)}(x)\SlDr\left[\SlDr\right]^{\dagger}e^{-\tau\SlDr\left[\SlDr\right]^{\dagger}} +\overline{W}^{(r)}(x)\left[\SlDr\right]^{\dagger}\SlDr e^{-\tau\left[\SlDr\right]^{\dagger}\SlDr} \Bigr) 
\notag \\
&= \mathop{\mathrm{Tr}}\Bigl(W^{(r)}(x)e^{-\delta\SlDr\left[\SlDr\right]^{\dagger}} +\overline{W}^{(r)}(x) e^{-\delta\left[\SlDr\right]^{\dagger}\SlDr} \Bigr) ,
\label{omega_div1}
\end{align}
where $W^{(r)}(x)$ and $\overline{W}^{(r)}(x)$ are defined as 
\begin{align}
W^{(r)}(x)&=\frac{d\Omega^{(r)}}{d r}\left[\Omega^{(r)}\right]^{-1}+\left(\frac{d\Omega^{(r)}}{d r}\left[\Omega^{(r)}\right]^{-1}\right)^{\dagger} ,\\
\overline{W}^{(r)}(x)&=\left[\overline{\Omega}^{(r)}\right]^{-1}\frac{d\overline{\Omega}^{(r)}}{d r}+\left(\left[\overline{\Omega}^{(r)}\right]^{-1}\frac{d\overline{\Omega}^{(r)}}{d r}\right)^{\dagger} .
\end{align}
To obtain Eq.~(\ref{omega_div1}), we used the cyclic property of trace, and
\begin{equation}
e^{-\tau\left[\SlDr\right]^{\dagger}\SlDr}\left[\SlDr\right]^{\dagger}=\left[\SlDr\right]^{\dagger}e^{-\tau\SlDr\left[\SlDr\right]^{\dagger}} .
\end{equation}
The small $\delta$ expansion of $d\omega(r)/dr$ reads
\begin{align}
\frac{d\omega(r)}{d r}&=
\frac{1}{4\pi \delta}\Tr\left(W^{(r)}(x)+\overline{W}^{(r)}(x)\right)
\notag\\
&\quad+\frac{1}{4\pi}\Tr\left(W^{(r)}(x)\alpha^{(r)}(x)+\overline{W}^{(r)}(x)\beta^{(r)}(x)\right)+{\rm O}(\delta) .
\label{omega_div2}
\end{align}
$\alpha^{(r)}$ and $\beta^{(r)}$ are the first Seeley coefficients in the small $\delta$ expansion of the heat kernels, and are defined as
\begin{align}
\langle x|e^{-\delta \SlDr\bigl(\SlDr\bigr)^{\dagger}}|x\rangle &= \frac{1}{4\pi\delta}\left(1+\delta \alpha^{(r)}(x)+{\rm O}(\delta^2)\right) , \\
\langle x|e^{-\delta \bigl(\SlDr\bigr)^{\dagger}\SlDr}|x\rangle &= \frac{1}{4\pi\delta}\left(1+\delta \beta^{(r)}(x)+{\rm O}(\delta^2)\right) .
\end{align}
The first Seeley coefficients are given as~\cite{Frishman:1992mr,Fujikawa:1979ay}
\begin{align}
\alpha^{(r)}&=\beta^{(r)}=\frac{\gamma^5}{2}\epsilon_{\mu\nu}\left(F^{(r)}_{\mu\nu}+G^{(r)}_{\mu\nu}\right) , \\
F^{(r)}_{\mu\nu}&\equiv-e(\partial_\mu a_\nu^{(r)}-\partial_\nu a_\mu^{(r)}) , \\
G^{(r)}_{\mu\nu}&\equiv\partial_\mu A_\nu^{(r)}-\partial_\nu A_\mu^{(r)}-i[A_\mu^{(r)},A_\nu^{(r)}].
\end{align}
We also have
\begin{equation}
W^{(r)}(x)=\overline{W}^{(r)}(x)=\left(\left[u^{(r)}\right]^{-1}\partial_r u^{(r)}+\partial_r v^{(r)}\left[v^{(r)}\right]^{-1}+\left[U^{(r)}\right]^{-1}\partial_r U^{(r)}+\partial_r V^{(r)}\left[V^{(r)}\right]^{-1}\right)\gamma^5 ,
\end{equation}
where we used $u^\dagger=v$ and $U^{\dagger}=V$.
Then, Eq.~(\ref{omega_div2}) becomes 
\begin{align}
\frac{d\omega(r)}{d r}&=
\frac{1}{2\pi}\int d^2x_\parallel\;\tr\left(\left[u^{(r)}\right]^{-1}\partial_ru^{(r)}+\partial_rv^{(r)}\left[v^{(r)}\right]^{-1}\right)\epsilon_{\mu\nu}F^{(r)}_{\mu\nu}\notag\\
&\quad+\frac{1}{2\pi}\int d^2x_\parallel\;\tr\left(\left[U^{(r)}\right]^{-1}\partial_rU^{(r)}+\partial_rV^{(r)}\left[V^{(r)}\right]^{-1}\right)\epsilon_{\mu\nu}G^{(r)}_{\mu\nu} +O(\delta).
\label{omega2}
\end{align}
The trace ``tr'' runs over the color space.
Since this expression is analytic in $\delta = 0$, we can take the limit of $\delta \to 0$.
Introducing gauge covariant quantities $g^{(r)}\equiv u^{(r)}v^{(r)}$ and $G^{(r)}\equiv U^{(r)}V^{(r)}$, we obtain the relation
\begin{align}
V^{(r)}\partial_-\left(\left[G^{(r)}\right]^{-1}\partial_+G^{(r)}\right)\left[V^{(r)}\right]^{-1}
&=i\partial_+A^{(r)}_--i\partial_-A^{(r)}_+ +\left[A^{(r)}_+,A^{(r)}_-\right]
\notag \\
&= \epsilon_{\mu\nu}G^{(r)}_{\mu\nu},
\label{Gr}
\end{align}
and a similar relation between $g^{(r)}$ and $F^{(r)}_{\mu\nu}$.
Here, we used $(\partial_-U_r)U_r^{-1}=-U_r\partial_-U_r^{-1}$, $(\partial_-V_r)V_r^{-1}=-V_r\partial_-V_r^{-1}$. 
We also obtain the relation
\begin{equation}
V^{(r)}\left[G^{(r)}\right]^{-1}\partial_rG^{(r)}\left[V^{(r)}\right]^{-1}=\left[U^{(r)}\right]^{-1}\partial_rU^{(r)}+\partial_rV^{(r)}\left[V^{(r)}\right]^{-1} .
\label{dotGr}
\end{equation}
and that of $g^{(r)}$.
From Eqs.~(\ref{Gr}) and~(\ref{dotGr}), and those of $g^{(r)}$,
Eq.~(\ref{omega2}) is reduced to the gauge invariant form
\begin{align}
\frac{d\omega(r)}{d r}
&=\frac{1}{2\pi}\int d^2x_\parallel\;\tr\left[\left[g^{(r)}\right]^{-1}\partial_rg^{(r)}\partial_-\left(\left[g^{(r)}\right]^{-1}\partial_+g^{(r)}\right)\right]
\notag \\
&\quad+\frac{1}{2\pi}\int d^2x_\parallel\;\tr\left[\left[G^{(r)}\right]^{-1}\partial_rG^{(r)}\partial_-\left(\left[G^{(r)}\right]^{-1}\partial_+G^{(r)}\right)\right]
\notag \\
&=
\frac{1}{2\pi}\int d^2x_\parallel\;\tr\left[\left[g^{(r)}\right]^{-1}\partial_rg^{(r)}\partial_\mu\left(\left[g^{(r)}\right]^{-1}\partial_\mu g^{(r)}\right)\right]
\notag \\
&\quad-\frac{i\epsilon_{\mu\nu}}{2\pi}\int d^2x\;\tr\left[\left(\left[g^{(r)}\right]^{-1}\partial_rg^{(r)}\right)\left(\left[g^{(r)}\right]^{-1}\partial_\mu g^{(r)}\right)\left(\left[g^{(r)}\right]^{-1}\partial_\nu g^{(r)}\right)\right]
\notag \\
&\quad+\frac{1}{2\pi}\int d^2x_\parallel\;\tr\left[\left[G^{(r)}\right]^{-1}\partial_rG^{(r)}\partial_{\mu}\left(\left[G^{(r)}\right]^{-1}\partial_\mu G^{(r)}\right)\right]
\notag \\
&\quad-\frac{i\epsilon_{\mu\nu}}{2\pi}\int d^2x_\parallel\;\tr\left[\left(\left[G^{(r)}\right]^{-1}\partial_rG^{(r)}\right)\left(\left[G^{(r)}\right]^{-1}\partial_\mu G^{(r)}\right)\left(\left[G^{(r)}\right]^{-1}\partial_\nu G^{(r)}\right)\right] .
\label{domega}
\end{align}
By integrating Eq.~(\ref{domega}) and dropping the surface terms, we obtain 
\begin{equation}
-W_{\rm E}[a_\mu,A_\mu] = \frac{1}{2}\int^1_0 dr \frac{d\omega(r)}{dr} =\Gamma_{\rm E}[uv]+\Gamma_{\rm E}[UV],
\end{equation}
where $\Gamma_{\rm E}[G]$ is the Euclidean version of the Wess-Zumino-Witten action.
By conducting analytic continuation, $-W_{\rm E}[a_\mu,A_\mu]\rightarrow iW[a_\mu,A_\mu]$, 
and $-\Gamma_{\rm E}[G]\rightarrow i\Gamma[G]$,
we have
\begin{equation}
W[a_\mu,A_\mu]= -\Gamma[uv]-\Gamma[UV]
\end{equation}
in the Minkowski space.

\end{document}